\documentclass[12pt]{article}
\usepackage{epsf}
\begin{document}

\begin{center}

{\large \bf EVENT GENERATOR BENCHMARKING FOR \\
PROTON RADIOGRAPHY APPLICATIONS}\\
\vspace{0.8cm}

{\bf Stepan G. Mashnik, Richard E. Prael, Arnold J. Sierk}\\
Los Alamos National Laboratory,  Los Alamos, NM 87545, USA\\

\vspace{0.8cm}
{\bf Konstantin K. Gudima}\\
Institute of Applied Physics\\   
Academy of Science of Moldova, Kishinev, MD-2028, Moldova\\

\vspace{0.8cm}
{\bf Nikolai V. Mokhov}\\
Fermi National Accelerator Laboratory, Batavia, IL 60510, USA\\

\vspace{0.8cm}
{\bf Abstract}
\end{center}
We have benchmarked the QGSM code and event generators of the MARS and LAHET3 codes
as potential candidates for high-energy programs  to be used 
in simulations for the Proton Radiography (PRad) Project.
We have compiled from the literature
experimental data on spectra of particles emitted
from proton-induced reactions at incident energies from 30 GeV
to 70 GeV on different nuclei and have  performed calculations 
for all reactions for which we found  data 
with these three codes without any modifications and
using only default parameters and standard inputs.
Our results (514 plots) show that all three codes describe reasonably 
most of the studied
reactions, though all of them should be further improved before becoming 
reliable tools for PRad.  
We present here our
conclusions concerning the relative roles of different reaction
mechanisms in the production of specific secondary particles. We comment 
on the strengths and weaknesses of QGSM, MARS, and LAHET3 
and suggest further improvements to these codes and to other models.

\newpage

{\noindent \bf Introduction}\\

The process of determining the feasibility of Proton Radiography (PRad)
[1-3]
as the radiographic probe for the Advanced Hydrotest Facility 
as well as its design and operation
require information about spectra of secondary particles produced by high 
energy protons interacting in the target and structural materials.
Reliable models and codes are needed to provide such data.
We studied the literature and chose three potential candidates for
high-energy codes  that may be used in simulations for PRad, namely 
the Quark-Gluon String Model (QGSM) 
as developed by Amelin, Gudima, and Toneev \cite{QGSM},
the MARS code by Mokhov {\em et al.} \cite{MARS},
and a version of the Los Alamos National Laboratory (LANL) transport code  
LAHET \cite{LAHET}, known as  LAHET3 \cite{LAHET3}.

The energy of the proton beam at PRad is supposed to be about 50 GeV.
Unfortunately, there are very few measurements of particle spectra
for proton-induced reactions exactly at 50 GeV or very close energies. 
In fact, we found only one published work at 50 GeV, namely spectra
of $\pi^-$ and $\pi^+$ measured at 159$^{\circ}$ from p(50 GeV) +
W published in Russian together with pion spectra for other energies 
and targets, in a 
Joint Institute for Nuclear Research (Dubna) Communication by
Belyaev {\it et al.} \cite{Belyaev89}. 

With only a few data available at 
50 GeV, we benchmarked QGSM,  MARS, and LAHET3
against measured spectra of particles emitted
from interaction of protons with energies $50 \pm 20$ GeV, {\it i.e.,} 
from 30 to 70 GeV, with all targets for which we found experimental 
data. Independently of how many spectra were measured in an experiment,
we performed calculations with the standard versions 
of QGSM, MARS, and LAHET3 without any modifications or adjustments, 
using only default parameters in the input of codes, and calculated 
double differential cross sections at 0, 4.75, 9, 13, 20, 45, 60, 90, and 
159 degrees, angle-integrated energy spectra, and mean multiplicities for
emission of n, p, d, t, $^3$He, $^4$He, $\pi^+$, $\pi^-$, $K^+$, $K^-$,
and  $\bar{p}$ for all cases listed below in Table 1.
The next Section presents a brief description of the benchmarked codes, 
followed
by results, discussion, and conclusions in the last two Sections.
\\

{\noindent \bf Benchmarked Codes}\\

{\it QGSM}: 
The core of the QGSM is built on a time-dependent version of the
intranuclear cascade model developed at Dubna to describe both particle-
and nuclei-induced reactions, often referred in the literature simply
as the Dubna intranuclear Cascade Model (DCM) (see \cite{Toneev83} 
and references therein).
The DCM models interactions of fast cascade particles (``participants")
with nucleon spectators of both the target and projectile nuclei and
includes interactions of two participants (cascade particles) as well.
It uses experimental cross sections (or those calculated by the Quark-Gluon 
String Model for energies above 4.5 GeV/nucleon) for these
elementary interactions to simulate angular and energy distributions
of cascade particles, also considering the Pauli exclusion
principle. When the cascade stage of a reaction is completed, QGSM uses the
coalescence model described in \cite{Toneev83}
to ``create" high-energy d, t, $^3$He, and $^4$He by
final state interactions among emitted cascade nucleons, already outside 
of the colliding nuclei.
After calculating the coalescence stage of a reaction, the QGSM
moves to the description of the last slow stages of the interaction,
namely to preequilibrium decay and evaporation, with a possible competition
of fission using the standard version of the Cascade Exciton Model (CEM)
\cite{CEM}. But if the residual nuclei have atomic numbers with  
$A \le 13$, QGSM uses the Fermi break-up model to calculate their further 
disintegration instead of using the preequilibrium and evaporation models.

{\it MARS}:
The MARS Monte-Carlo code system, being developed over 29~years,
allows fast and reliable inclusive and exclusive simulation of 
three-dimensional hadronic and electromagnetic cascades in shielding, 
accelerator and detector components in the energy range from a fraction 
of an electron-volt up to about 100~TeV~\cite{MARS}. It is under 
continuous development. The reliable performance of the code has been 
demonstrated in numerous applications at Fermilab, CERN, KEK and other 
centers as well as in special benchmarking studies.
Description of elastic and inelastic $hN$, $hA$, $\gamma A$ and $\nu A$ 
cross sections is based on the newest compilations and 
parameterizations~\cite{MARS98}.
At high energies (5~GeV$<$E$<$100~TeV), $\sigma_{tot}$, $\sigma_{in}$, 
$\sigma_{prod}$ and $\sigma_{el}$ are calculated in the framework of the 
Glauber multiple scattering theory with the $\sigma_{hN}$ as an input.
The nucleon density distribution in nuclei is represented as the 
symmetrized Fermi function with the parameters of~\cite{Alk78} for medium and 
heavy nuclei ($Z>10$) and the ones of~\cite{Bur76} for $Z<10$. 
Modern evaluated nuclear data as well as fitting formulae are used to simulate
hadron-nucleus elastic scattering. For protons, nuclear, Coulomb elastic 
scattering, and their interference is taken into account.
At E$>$5~GeV, a simple analytical description used in the code for both 
coherent and incoherent components of $d\sigma /dt$ is quite consistent with
experiment.
A version of the Cascade-Exciton Model of nuclear reactions~\cite{CEM}
as realized in the code CEM95~\cite{CEM95} and
containing also several recent refinements~\cite{CEM98} is
now implemented in the 1998 version of MARS \cite{MARS98}
 as default for 1-10~MeV $<$ E $<$ 3-5~GeV.
A set of phenomenological models, as described in 
Ref.~\cite{MARS,book89,mokstr98},
is used for inclusive production of secondary particles in 
$hA$, $dA$, $\gamma A$ and $\nu A$ interactions at projectile energies 
from 5 GeV to 100 TeV.
The 2001 version \cite{MARS98} of the MARS code was employed in the present 
benchmark.

{\it LAHET3}: 
LAHET is a Monte-Carlo code for the transport and interaction of nucleons,
pions, muons, light ions, and antinucleons in complex geometry~\cite{LAHET};
it may also be used without particle transport to generate particle production
cross sections. LAHET allows one to choose one
of several options for the
Intra-Nuclear Cascade (INC) and fission models to be employed in
calculations; it is widely used and well known in the applied nuclear physics
community; therefore, we do not describe it here (a comprehensive description 
of LAHET may be found in \cite{LAHET} and references therein).
The version of LAHET realized in the code LAHET3 \cite{LAHET3}
uses a version of the code FLUKA, known in the literature as
FLUKA96 \cite{FLUKA96} to describe the first, INC stage
of reactions, and its own Multistage Preequilibrium Model (MPM) \cite{MPM}
to describe the following intermediate preequilibrium stage, followed
by evaporation/fission slow processes (or by the Fermi break-up
model after the cascade instead of preequilibrium and evaporation/fission, 
if the residual nuclei have atomic numbers with  $A \le 13$
and for $ 14 \le A \le 20$ with excitation energy below 44 MeV), as 
described in \cite{LAHET,LAHET3}. We mention again that only
the high-energy event generator from FLUKA96 is employed here, as 
implemented in LAHET3; the default preequilibrium, evaporation and Fermi 
break-up models of LAHET3 are used for low energy nucleon and complex 
particle emission. More details and further references on LAHET3 with
FLUKA96 can be found in \cite{FLUKA96R}.
\\

{\noindent \bf Results and Discussion}\\

Table 1 lists the cases we calculated with QGSM, MARS, and LAHET3,
and provides references to experimental works where at least one
spectrum of a secondary particle (from the ones listed in Introduction)
was measured. 
A detailed report of the study containing 514 plots with
spectra and multiplicities of secondary particles from reactions listed
in Tab.\ 1 is now in preparation. Here, we present only our main conclusions
and several

\begin{center}

{\bf Table 1. Proton energy and target list covered by
the present benchmark}\\

\vspace*{5mm}
\begin{tabular}{|c|c|c|}
\hline \hline
 $T_p$ (GeV) & Nuclei & Measurements \\
\hline
30& $^9$Be, $^{27}$Al & [8, 22--24] \\
47& $^{12}$C & [8, 25] \\
50& $^{184}$W & [8, 25] \\
51& $^9$Be, $^{48}$Ti & [8, 25] \\
53& $^{27}$Al & [8, 25] \\
54& $^{96}$Mo & [8, 25] \\
70& $^{12}$C, $^{27}$Al, $^{64}$Cu, $^{118}$Sn, $^{208}$Pb & [26--30]\\
\hline \hline
\end{tabular}

\end{center}

{\noindent
 examples of results from the study.}

Our analyses have shown that all three codes tested here describe
reasonably most of the secondary particle spectra. As a rule the
higher the incident proton energy, the better the calculated spectra
agree with experimental data. Several reaction mechanisms participate
in the production of secondary nucleons and complex particles. 
These mechanisms are: 1) Fast INC processes; 2) preequilibrium emission
from residual nuclei after the cascade stage of reactions; 3) evaporation
of particles from compound nuclei after the preequilibrium stage, or
from fission fragments, if the compound nucleus was heavy enough to
undergo fission; 4) Fermi break-up of light excited nuclei formed
after the cascade stage of reactions; 5) coalescence of complex particles
by final state interactions among emitted cascade nucleons; 6) fast 
complex particle emission via knock out and pick up processes;
7) Multifragmentation of highly-excited residual nuclei after the INC. 
Their
relative roles change significantly with the changing atomic mass
number of the targets, and are different for different energies and
angles of emission of secondary particles. 
Different codes describe these spectra better,
worse, or do not describe them at all, depending of how these reaction
mechanisms are (or are not) implemented into a specific code.

As an example, Fig.\ 1 shows spectra of p, d, t, and $\pi^-$ emitted
at 9.17 deg from the reaction p(70 GeV) + $^{208}$Pb. Results for
other reactions at 70 GeV are similar. One can see that all three
codes describe the proton spectra well. The agreement for the
pion spectra is not so good but is still reasonable, with some underestimation
of the high-energy tails of spectra by QGSM and some overestimation
by MARS. Note that as the angle of pion emission changes the situation 
is reverses: we observe that most of the high-energy
tails of pion spectra at 159 deg, and to a lesser extent at 90 deg,
are over-predicted by LAHET3 and underestimated by MARS. 

The situation with the deuteron and tritium spectra
is quite different. 
We see that deuterons with momentum of up to about 
15 GeV/c and tritium with momenta up to 19 GeV/c
are emitted and measured in this particular reaction. 
Utilizing the coalescence mechanism for complex particle emission, QGSM
is able to describe high-energy deuteron production, and agrees well with
the measurement. LAHET3 does not consider
the coalescence of complex particles and therefore describes emission of
only evaporative and preequilibrium deuterons with momenta not 
higher than 1 GeV/c. MARS does not consider emission of complex
particles at such high incident proton energies, therefore no
d and t spectra by MARS are shown in Fig.\ 1.

For tritium, the situation is worse since 
LAHET3, as is the case of deuterons, describes only preequilibrium 
emission and evaporation of tritons
with momenta not higher than 1 GeV/c and
QGSM, even taking into account coalescence of tritium,
describes emission of t from this reaction up to only 2.5 GeV/c 
while the experimental spectrum of t extends to 19 GeV/c. 
This deficiency can be understood by considering the
coalescence mechanism: It is much more
probable to emit two cascade
nucleons with very similar momenta that can coalesce into a deuteron 
than to get three INC nucleons with very similar momenta that
can coalesce into a triton. The experimental values of high-energy
triton spectra are several orders of magnitude below the corresponding 
values of the deuteron spectra, and the statistics of our present
QGSM simulation could be simply
too small to get such
high-energy tritium via coalescence.
There is also a possibility that knock out processes of preformed
clusters (or fluctuations of nuclear density, leading to ``fluctons")
by bombarding protons are seen in these experimental
d and t spectra, but are not taken into account by any of the tested 
codes, 
providing the observed difference in the t spectrum and less pronounced,
in the d spectrum. A third possible mechanism of 
complex particle emission with greater than 1 GeV/c momenta
would be multifragmentation of highly-excited residual
nuclei after the INC. This mechanism is not taken into account by 
any of the tested codes and we cannot estimate its contributation.

Fig.\ 2 shows examples of $\pi^+$ spectra at 159 deg from 51 GeV proton
collisions with $^9$Be and $^{48}$Ti.  As already mentioned above for
$\pi^-$, we see 
that LAHET3 overestimates the high-energy tails of pion spectra and MARS
underestimates them a little. Similar results were obtained for other
targets and incident proton energies.

Fig.\ 3 shows an example of how calculated proton spectra depend on the
angle of emission, for the reaction p(30 GeV) + $^9$Be. We see that
at 30 GeV, the agreement of calculated proton spectra with the data is 
not so good as we have in Fig.\ 1 for 70 GeV. The shapes
and absolute values of proton spectra predicted by different codes
depend significantly on the angle of detection, as does the
agreement with the data. Similar results were obtained for other
secondary particles and for other targets and incident energies.

\begin{figure}
\vspace{-7.cm}
\centerline{\hspace{-2cm} \epsfxsize 21cm \epsffile{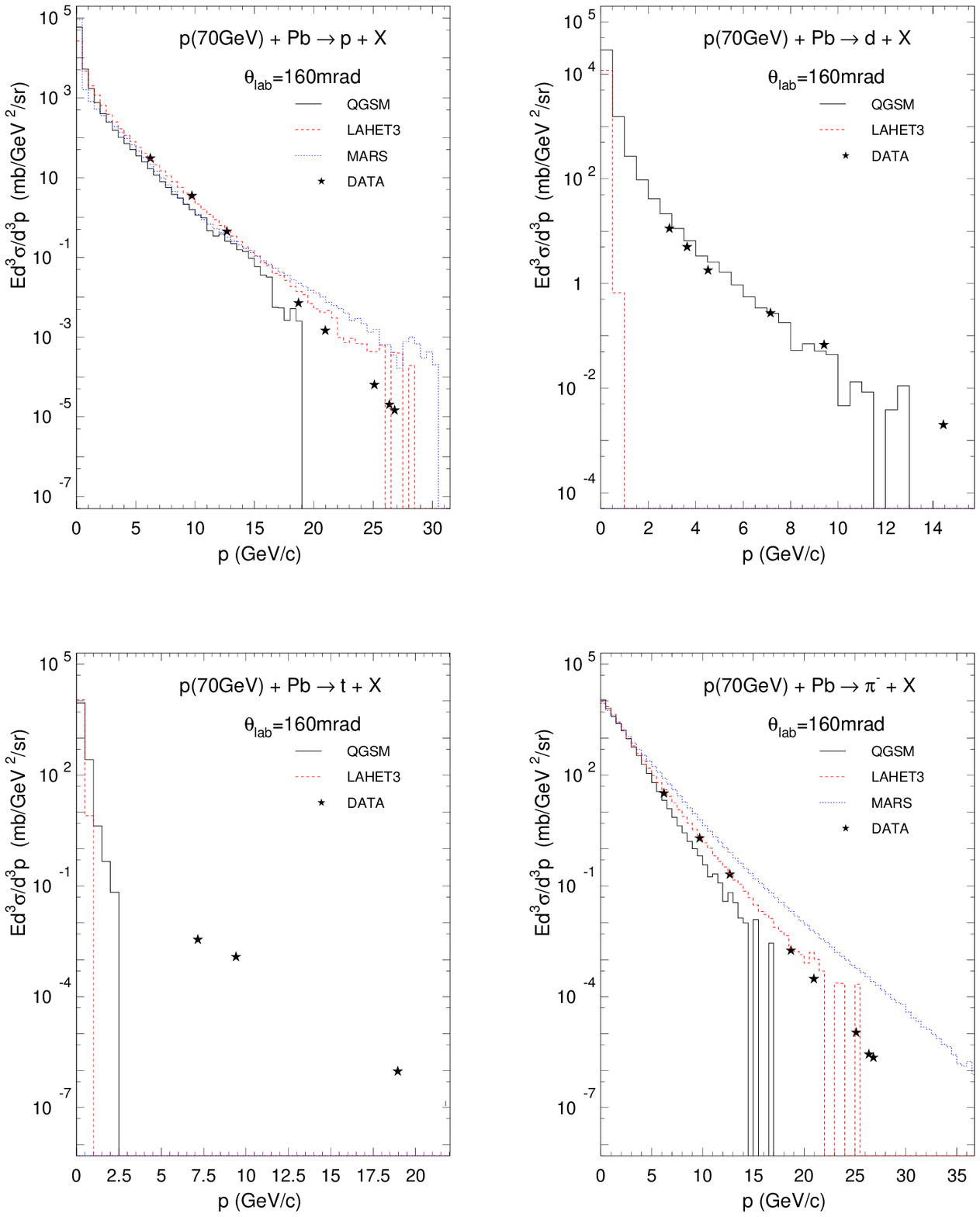}} 
\vspace{-5.8cm}
{\bf Figure 1.}
Invariant cross sections $E d^3 \sigma /d^3 p$ for forward production
of p, d, t, and $\pi^-$ at 160 mrad (9.17 deg) as functions of
particle momentum $p$ from 70 GeV protons on $^{208}$Pb.
Experimental data for p and $\pi^-$ are from Tab.\ 1 of 
Ref.\ \cite{Abramov85} and for d and t, from 
Ref.\ \cite{Abramov87}. Calculations by QGSM, LAHET3, and MARS
are shown as indicated in the 
legends.
\end{figure}

\newpage
\begin{figure}
\vspace{-5.cm}
\centerline{\hspace{-2cm} \epsfxsize 21cm \epsffile{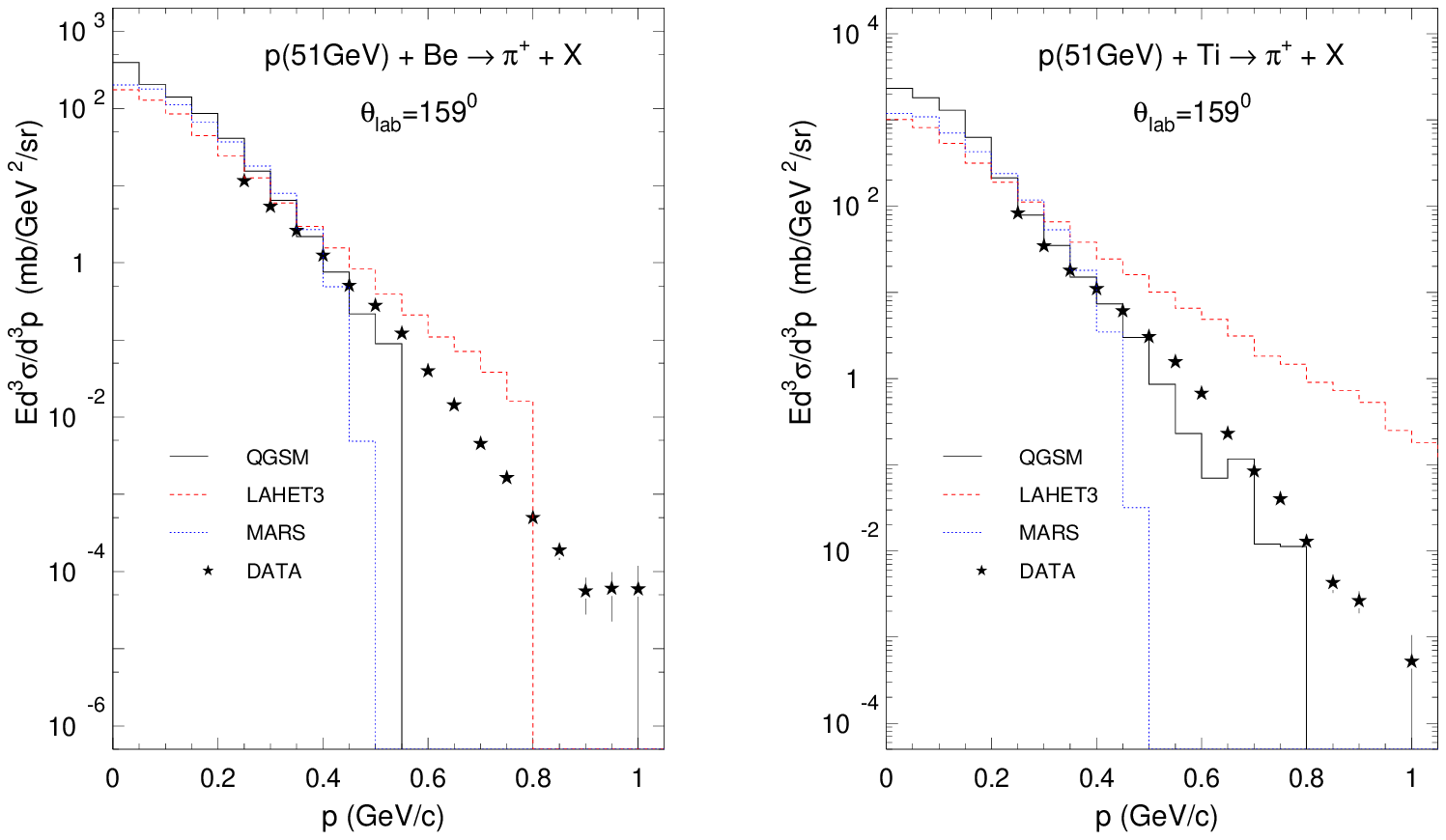}} 
\vspace{-16.5cm}
{\bf Figure 2.}
Invariant cross sections $E d^3 \sigma /d^3 p$ for production
of $\pi^+$ at 159 deg as functions of
pion momentum $p$ from 51 GeV protons on $^{9}$Be and $^{48}$Ti.
Experimental data are by Belyaev {\em et al.} 
\cite{Belyaev89}.
Calculations by QGSM, LAHET3, and MARS
are shown as indicated in legends.
\end{figure}

{\noindent
}

\begin{figure}
\vspace{-7.cm}
\centerline{\hspace{-2cm} \epsfxsize 21cm \epsffile{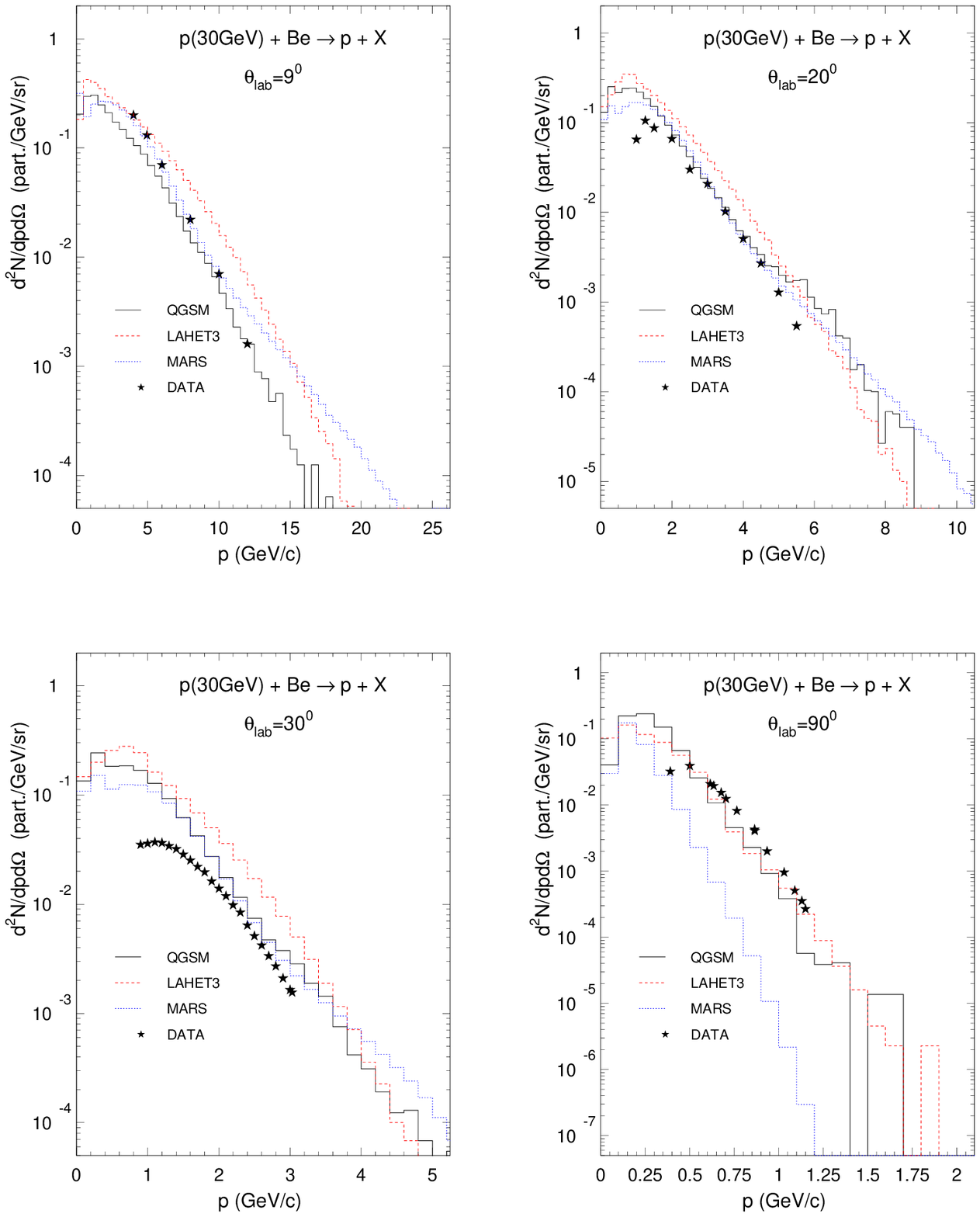}} 
\vspace{-5.8cm}
{\bf Figure 3.}
Momentum spectra of secondary protons from 30 GeV protons on Be.
Experimental data at 9 and 20 degrees are taken from Fig.\ 1 of
Baker {\em et al.} \cite{Baker}; at 30 degrees, from Fig.\ 5 of
Ref. \cite{Schwarzschild}; and at 90 degrees, from Fig.\ 2
of Ref. \cite{Fitch}. Calculations by QGSM, LAHET3, and MARS
are shown as indicated in legends.

\end{figure}

\begin{figure}
\vspace{-7.cm}
\centerline{\hspace{-2cm} \epsfxsize 21cm \epsffile{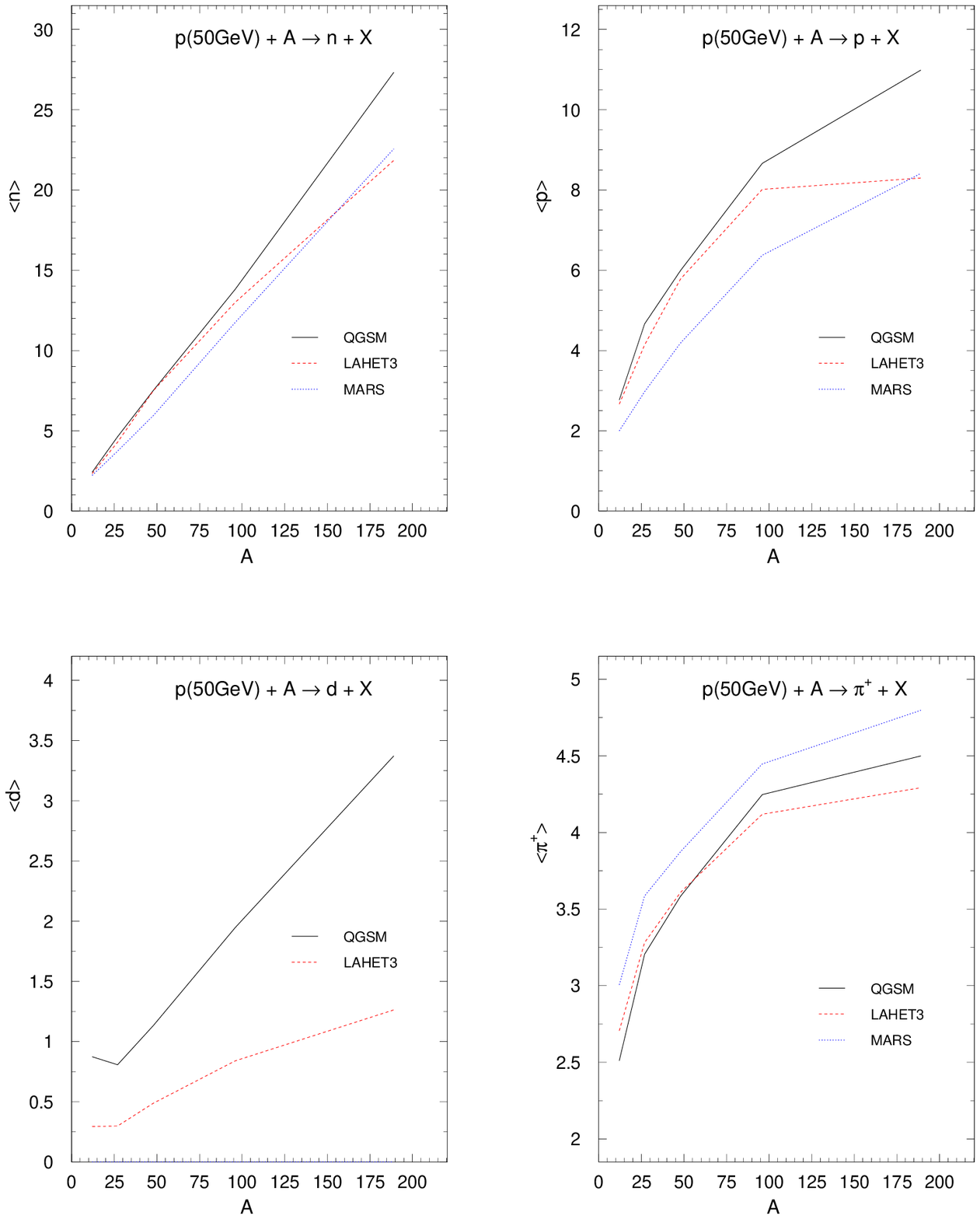}}
\vspace{-6.8cm}
{\bf Figure 4.}
Predicted by QGSM, LAHET3, and MARS
mean multiplicities of secondary n, p, d, and $\pi^+$ emitted from
50 GeV proton-induced reactions as functions the mass number of
targets. Note that actual energies of incident protons in our 
calculations were 47 GeV for $^{12}$C, 50 GeV for $^{184}$W,
51 GeV for $^{9}$Be and $^{48}$Ti, 53 GeV for $^{27}$Al, and
54 GeV for $^{96}$Mo. (MARS does not calculate production of deuterons
at these incident energies.)

\end{figure}

{\noindent
}
Fig.\ 4 shows an example of mean multiplicity of secondary n, p, d, and
$\pi^+$ predicted by the tested codes
for interaction of protons of about 50 GeV with
different nuclei as functions of the mass number of targets. We see
that predicted
particle multiplicities
 differ
significantly from each other, and the differences increase with
increasing mass number of the target. 
The observed differences point to a quite significant
difference in the treatment by the codes
of the cascade stage of reactions (pions
are emitted only at the cascade stage of reactions) and of the
subsequent preequilibrium, evaporation, and Fermi break-up stages as
well (we recall that at these incident energies MARS uses
its own approximations for the total particle spectra without considering 
separately contributions from different mechanisms of nuclear reactions).
These differences indicate that further experimental data are necessary
at these incident proton energies and further development and improvement
of the codes is required.\\

{\noindent \bf Further Work} \\

Our study shows that all three 
codes describe reasonably well many of the secondary particle spectra
analyzed here, though all of them should 
be further improved before becoming reliable tools for PRad.  

For instance, 
we find that QGSM has some problems in a correct description of
several pion spectra and does not describe sufficiently the high-energy
tails of measured t and $^3$He spectra. Nevertheless, QGSM is the only
code tested here that accounts for coalescence of complex particles from
cascade nucleons and provides production of high-energy complex particles.

MARS overestimates the high-energy tails of some pion, kaon, and proton
spectra at small angles (4.75$^{\circ}$, 9$^{\circ}$, and 13$^{\circ}$)
and underestimates them a little at large angles (90$^{\circ}$ and
159$^{\circ}$). At these incident energies, MARS does not calculate
complex particle production. However, MARS has 
one significant advantage in comparison with the two other codes: It is several
orders of magnitude faster and requires almost no computing time,
providing meanwhile reliable results for many 
applications.

LAHET3 overestimates the high-energy tails of practically all pion spectra
at 159$^{\circ}$ and some nucleon and
complex particle spectra in the preequilibrium energy region. It does not 
consider coalescence of complex particles and does not describe production
of high-energy complex particles.

We observe also big differences between predicted 
high-energy tails of both neutron and proton spectra at 0$^{\circ}$
and for the mean multiplicities of almost all secondary particles,
though no experimental data for these quantities are available at present
for the reactions studied here.

We note that many of the problems we observe in our study for particular codes
have already been solved, since all benchmarked event generators
are under continuous development and
improvement and all of them have been further improved in comparison
with the versions we use in this study. 

On the basis
of QGSM, we have developed the Los Alamos version of the Quark-Gluon String 
Model code, LAQGSM \cite{LAQGSM}.
LAQGSM differs from QGSM by replacing the preequilibrium and
evaporation parts  of QGSM described according to the standard CEM 
\cite{CEM} with the new physics from CEM2k \cite{CEM2k}
and has a number of improvements 
and refinements in the cascade and Fermi break-up models. 
Originally, both QGSM and LAQGSM were not able to describe fission reactions 
and production of light fragments heavier than $^4$He, as they had neither 
a high-energy-fission nor a fragmentation model. 
Recently, we addressed these problems \cite{CEM2kGEM2}
by further improving CEM2k and LAQGSM
and by merging them with the Generalized Evaporation Model
code GEM2 developed by Furihata \cite{GEM2}. The improved LAQGSM+GEM2 code
describes both spectra of secondary particles and yields of produced
nuclides much better than QGSM does; exemplary results
by LAQGSM and further references may be found in \cite{COSPAR02}.

The MARS code system is being continuously developed and improved.
For instance, a new version of the code, MARS14(2002) was
completed after we started the present work. It contains a large
number of improvements and refinements and provides better
results in comparison with the version used here.
Recently, the authors of MARS started to develop new and better
approximations for the double differential cross sections of 
inelastic $hN$ and $hA$ interactions. The new systematics allow
to solve the mentioned above problems with the pion, kaon, and 
proton spectra at forward and large angles and describe the experimental
data much better.
 
The FLUKA code has also been updated very significantly
(see {\it e.g.,} \cite{FLUKA} and references therein) 
since the version FLUKA96 was incorporated into LAHET3 as used 
here; no updated version is yet incorporated into LAHET.

Our study points to the importance of taking into account 
coalescence in high-energy complex-particles production.
We find it appropriate and easy to implement these
processes into MARS and LAHET, as well as into other codes that do
not now consider coalescence.

We think that at such high incident energies, multifragmentation of
highly-excited heavy nuclei may also be of significance and
should be taken into account in these 
event generators and in other codes.

\begin{center}
\vspace{0.8cm}
{\it Acknowledgment}
\end{center}

\noindent

The study was supported by the U. S. Department of Energy and by the 
Moldovan-U.~S. Bilateral Grants Program, CRDF Project MP2-3025.



\begin{thebibliography}{99}

\bibitem{Morris00}      
Christopher L. Morris,
``Proton Radiography for an Advanced Hybrotest Facility,''
Los Alamos National Report LA-UR-00-5716, 2000;
http://lib-www.lanl.gov/la-pubs/00357005.pdf.

\bibitem{Ziock98}      
H.-J. Ziock {\em et al.}, 
``The Proton Radiography Concept,''
Los Alamos National Report LA-UR-98-1368, 1998;
http://lib-www.lanl.gov/la-pubs/00460235.pdf;\\
J. F. Amann {\em et al.}, 
``High-Energy Test of Proton Radiography Concepts,''
Los Alamos National Report LA-UR-97-1520, 1997;
http://lib-www.lanl.gov/cgi-bin/getfile?00366878.pdf.

\bibitem{King99}      
N. S. P. King {\em et al.}, 
``An 800-MeV Proton Radiography Facility for Dynamic Experiments,''
{\it Nucl. Instr. Meth. } {\bf A424} (1999) 84--91. 

\bibitem{QGSM}         
N.~S.~Amelin, K.~K.~Gudima, and V.~D.~Toneev,
``Ultrarelativistic Nucleus-Nucleus Collisions in a Dynamical Model of
Independent Quark-Gluon Strings,"
{\it Sov.~J.~Nucl.~Phys.}~{\bf 51} (1990) 1093--1101
[{\it Yad.~Fiz.}~{\bf 51} (1990) 1730--1743];
N.~S.~Amelin, K.~K.~Gudima, and V.~D.~Toneev,
``Further Development of the Model of Quark-Gluon Strings for the
Description of High-Energy Collisions with a Target Nucleus,"
{\it Sov.~J.~Nucl.~Phys.}~{\bf 52} (1990) 172--178
[{\it Yad.~Fiz.}~{\bf 52} (1990) 272--282];
N.~S.~Amelin,
``Simulation of Nuclear Collisions at High Energy in the Framework of 
the Quark-Gluon String Model,"
Joint Institute for Nuclear Research Report JINR-86-802, Dubna (1986).

\bibitem{MARS}        
N.~V.~Mokhov,
``The MARS Code System User's Guide,"
Fermilab-FN-628 (1995);
N.~V.~Mokhov and O.~E.~Krivosheev,
``MARS Code Status,"
Fermilab-Conf-00/181 (2000);
http://www-ap.fnal.gov/MARS/.

\bibitem{LAHET}      
R.~E.~Prael and H.~Lichtenstein, 
``User Guide to LCS: The LAHET Code System," 
LANL Report No.~LA-UR-89-3014, Los Alamos (1989); \\
http://www-xdiv.lanl.gov/XTM/lcs/lahet-doc.html.

\bibitem{LAHET3}      
R. E. Prael, ``Release of LAHET$^{TM}$ 
Version 3.00,'' Los Alamos National Laboratory
Research Note XCI-RN 98-10 (U), January 14, 1998; LA-UR-00-2116;
R. E. Prael, ``Release Notes for LAHET Code System with LAHET$^{TM}$ 
Version 3.16,'' LANL Report LA-UR-01-1655, Los Alamos, 2001;
http://lib-www.lanl.gov/la-pubs/00795117.pdf.

\bibitem{Belyaev89}         
I. M. Belyaev {\em et al.}, 
``The Cross Sections of $\pi^+$- and $\pi^-$ Production at an Angle 159$^\circ$
l.s. in Proton-Nuclear Interactions at the Energy of the Incident Protons
from 15 to 65 GeV,''
JINR Communication P1-89-112, Dubna, Russia (1989);
part of the Be data from this report are available
from the HEPDATA: REACTION DATA Database at the web page
http://cpt19.dur.ac.uk/hepdata/reac2.html.

\bibitem{Toneev83}         
V.~D.~Toneev and K.~K.~Gudima,
``Particle Emission in Light and Heavy-Ion Reactions,"
{\it Nucl.~Phys.}~{\bf A400} (1983) 173c--190c.

\bibitem{CEM} 
K. K. Gudima, S. G. Mashnik, and V. D. Toneev,
``Cascade-Exciton Model of Nuclear Reactions,"
{\it Nucl. Phys.} {\bf A401} (1983) 329--361. 

\bibitem{MARS98}        
N. V. Mokhov, S. I. Striganov, A. Van Ginneken,
S. G. Mashnik, A. J. Sierk, and J. Ranft,
``MARS Code Development,''
Fermilab-Conf-98/379 (1998); LANL Report LA-UR-98-5716 (1998);
Eprint: {\bf nucl-th/9812038};
{\em Proc. Fourth Int. Workshop on Simulating Accelerator Radiation
Environments (SARE-4), Hyatt Regency, Knoxville, TN, September 13--16, 1998},
edited by T. A. Gabriel,
ORNL (1999), pp. 87--99.

\bibitem{Alk78} 
G.~D.~Alkhazov, S. L. Belostotsky, and A. A. Vorobyov,
``Scattering of 1 GeV Protons on Nuclei," 
{\it Phys.~Rep.} {\bf 42} (1978) 89--144;
H. de Veries, C. W. de Jager, and C. de Vries,
``Nuclear Charge-Density-Distribution Parameters from Elastic
Electron Scattering,"
{\it Atomic Data and Nuclear Data Tables} {\bf 36} (1997) 495--536.

\bibitem{Bur76} 
V.~V.~Burov, D. N. Kadrev, V.~K.~Lukianov, and Yu.~S.~Pol, 
``Analysis of Charge-Density Distributions in Nuclei,"
{\it Phys. At. Nucl.} {\bf 61}
(1998) 525--532
[{\it Yad. Fiz.} {\bf 61}
 (1998) 595--602].

\bibitem{CEM95}
S.\ G.\ Mashnik, ``User Manual for the Code CEM95," JINR, Dubna (1995);
OECD NEA Data Bank, Paris, France (1995); 
http://www.nea.fr/abs/html/iaea1247.html;
RSIC-PSR-357, Oak Ridge, 1995.

\bibitem{CEM98}                 
S.~G.~Mashnik and A.~J.~Sierk, 
``Improved Cascade-Exciton Model of Nuclear Reactions'', 
Eprint: {\bf nucl-th/9812069};
{\em Proc. Fourth Int. Workshop on Simulating Accelerator Radiation
Environments (SARE-4), Hyatt Regency, Knoxville, TN, September 13--16, 1998},
edited by T. A. Gabriel,
ORNL (1999), pp. 29--51.

\bibitem{book89} 
A.~N.~Kalinovskii, N.~V.~Mokhov, Yu.~P.~Nikitin,
{\it Passage of High-Energy Particles through Matter},
AIP, NY, 1989.

\bibitem{mokstr98} 
N.~V.~Mokhov and S.~I.~Striganov,
``Model for Pion Production in Proton-Nucleus Interactions'',
Proc.~of the Workshop on Physics at the First Muon Collider,
Fermilab, November 1997, AIP Conf.~Proc.~No.~435, pp.~453--459;
Fermilab-Conf-98/053, 1998.

\bibitem{dpmjet} 
J.~Ranft, 
``Dual Parton Model at Cosmic Ray Energies,"
{\em Phys.~Rev.} {\bf D51}, (1995) 64--85;
Gran Sasso Report INFN/AE-97/45, 1997.

\bibitem{FLUKA96}      
A.~Fasso, A.~Ferrari, J.~Ranft, and P.~R.~Sala,
``An Update About FLUKA,"
{\em Proc.~of the Second Workshop on Simulating Accelerator
Radiation Environments (SARE-2), CERN, Geneva, October 9--11, 1996,}
edited by G.~R.~Stevenson, Geneva, Switzerland (1997), pp.~158--170;
more references 
and many details on FLUKA may be found at the Web page
http://pcfluka.mi.infn.it/.

\bibitem{MPM}    
R.~E.~Prael and M.~Bozoian,
``Adaptation of the Multistage Preequilibrium Model for the
Monte-Carlo Method (I),"
LANL Report LA-UR-88-3238, Los Alamos (1988).

\bibitem{FLUKA96R}        
R. E. Prael, Al. Ferrari, R. K. Tripathi, and A. Polanski,
``Comparison of Nucleon Cross Section Parametrization Methods for
Medium and High Energies,"
LANL Report LA-UR-98-5813 (1998);
http://www-xdiv.lanl.gov/XCI/PEOPLE/rep/;
{\em Proc. Fourth Int. Workshop on Simulating Accelerator Radiation
Environments (SARE-4), Hyatt Regency, Knoxville, TN, September 13--16, 1998},
edited by T. A. Gabriel,
ORNL (1999), pp. 171--181.


\bibitem{Baker}         
W. F. Baker {\em et al.}, 
``Particle Production by 10--30 BeV Protons Incident on Al and Be,''
{\it Phys. Rev. Let.} {\bf 7}
(1961) 101--104.

\bibitem{Schwarzschild}         
A. Schwarzschild and {\v C}. Zupan{\v c}i{\v c},
``Production of Tritons, Deuterons, Nucleons, and Mesons by 30-GeV
Protons on Al, Be, and Fe Targets,''
{\it Phys. Rev.} {\bf 129}
(1962) 854--862.

\bibitem{Fitch}         
V. L. Fitch, S. L. Meyer, and P. A. Pirou\'e,
``Particle Production at Large Angles by 30- and 33-BeV Protons Incident
on Aluminum and Beryllium,''
{\it Phys. Rev.} {\bf 126}
(1962) 1849--1851.

\bibitem{Gavrishchuk91}         
O. P. Gavrishchuk {\em et al.},
``Charged Pion Backward Production in 15--65 GeV Proton-Nucleus Collisions,''
 Nucl. Phys. {\bf A523} (1991) 589--596;
the Be data from this paper are available
from the HEPDATA: REACTION DATA Database at the web page
http://cpt19.dur.ac.uk/hepdata/reac2.html.

\bibitem{Barkov82}         
L. M. Barkov, {\em et al.},
``Production of Low-Energy Hadrons at Zero Angle in Proton-Nucleus Collisions
at Energy 70 GeV,''
{\it Sov. J. Nucl. Phys.} {\bf 35}
(1982) 694--698
[{\it Yad. Fiz.} {\bf 35}
 (1982) 1186--1193].

\bibitem{Barkov83}         
L. M. Barkov, {\em et al.},
``Measurement of the Cross Sections for Production of Hadrons with Momentum
up to 2 GeV/c in Proton-Nucleus Collisions at Energy 70 GeV,''
{\it Sov. J. Nucl. Phys.} {\bf 37}
(1983) 732--737
[{\it Yad. Fiz.} {\bf 37}
 (1983) 1232--1240].

\bibitem{Abramov84}         
V. V. Abramov {\em et al.},
``High $P_{\perp}$ Hadron Production off Nuclei at 70 GeV,''
{\it Z. Phys.} {\bf C24} (1984) 205--215.

\bibitem{Abramov85}         
V. V. Abramov {\em et al.},
``Production of hadrons with Large $P_{\bot}$ in Nuclei at 70 GeV,''
{\it Sov. J. Nucl. Phys.} {\bf 41} (1985) 227--236
[{\it Yad. Fiz.} {\bf 41} (1985) 357--370]; 
Preprint IFVE-84-26, Serpukhov (1984);
tabulated values are
available in the HEPDATA: REACTION DATA Database at the web page
http://cpt19.dur.ac.uk/hepdata/reac2.html.

\bibitem{Abramov87}         
V. V. Abramov {\em et al.},
``Production of Deuterons and Antideuterons with Large $p_{\perp}$ in $pp$
and $p A$ Collisions at 70 GeV,''
{\it Sov. J. Nucl. Phys.} {\bf 45}
 (1987) 845--851
[{\it Yad. Fiz.} {\bf 45}
 (1987) 1362--1372]; 
Preprint IFVE-86-56, Serpukhov (1986);
tabulated values are
available in the HEPDATA: REACTION DATA Database at the web page
http://cpt19.dur.ac.uk/hepdata/reac2.html.

\bibitem{LAQGSM}
K. K. Gudima, S. G. Mashnik, and A. J. Sierk,
``User Manual for the Code LAQGSM,"
Los Alamos National Report LA-UR-01-6804, 2001.

\bibitem{CEM2k}
S. G. Mashnik and A. J. Sierk,
``CEM2k---Recent Developments in CEM," 
{\it Proc. AccApp00 (Washington DC, USA)}, pp. 328--341,
La Grange Park, IL, USA,  2001; Eprint: {\bf nucl-th/0011064}; see also
Mashnik S. G., and A. J. Sierk,
``Recent Developments of the Cascade-Exciton Model of Nuclear Reactions,
{\it Proc. ND2001 (Tsukuba, Japan)},  
{\it J. Nucl. Sci. Techn.}, {\bf Supplement 2}, 720--725, 2002;
Eprint: {\bf nucl-th/0208074}.

\bibitem{CEM2kGEM2}
S. G. Mashnik, K. K. Gudima, and A. J. Sierk,
``Merging the CEM2k and LAQGSM Codes with GEM2 to Describe
Fission and Light-Fragment Production,"
{\em  Proc. SATIF-6 (SLAC, USA)}; LANL Report LA-UR-02-0608, Los Alamos, 2002;
S. G. Mashnik, A. J. Sierk, and K. K. Gudima,
``Complex-Particle and Light-Fragment Emission in the Cascade-Exciton Model
of Nuclear Reactions," 
{\em Proc. RPSD 2002 (Santa Fe, NM)};  
LANL Report LA-UR-02-5185, Los Alamos 2002; Eprint: {\bf nucl-th/0208048}.

\bibitem{GEM2}
S. Furihata,
``Statistical Analysis of Light Fragment Production from Medium Energy
Proton-Induced Reactions,"
{\it Nucl.\ Instr.\ Meth.\ }{\bf B171} (2000) 252--258;
S. Furihata,
``The Gem Code Version 2 Users Manual,"
Mitsubishi Research Institute, Inc., Tokyo, Japan, 2001.

\bibitem{COSPAR02}
S. G. Mashnik, K. K. Gudima, I. V. Moskalenko, R. E. Prael, and A. J. Sierk, 
``CEM2k and LAQGSM as Event Generators for Space-Radiation-Shielding
and Cosmic-Ray-Propagation Applications,''
Proc. Second World Space Congress, COSPAR 2002, Houston,
TX, USA, October 10--19, 2002;
LA-UR-02-6558, Los Alamos (2002);
Eprint: {\bf nucl-th/0210065};
to be published in the journal {\em Advances in Space Research}.

\bibitem{FLUKA}
A. Fasso, A. Ferrari, J. Ranft, and P. R. Sala,
``FLUKA: Status and Prospective for Hadronic Applications,"
invited talk in the {\it Proc. Monte-Carlo 2000 Conf., Lisbon, October 23--26,
2000}, A. Kling, F. Barao, M. Nakagawa, L. Tavora, P. Vaz eds.,
Springer-Verlag, Berlin pp. 955--960 (2001);
more references and many details on FLUKA nay be found on the Web page
http://www.fluka.org/.

\end{thebibliography}
\end{document}